# Fragile charge order in the non-superconducting ground state of the underdoped high temperature superconductors

B. S. Tan[1], N. Harrison[2], Z. Zhu[2], F. F. Balakirev[2], B. J. Ramshaw[2], A. Srivastava[1], S. A. Sabok[3], B. Dabrowski[3], G. G. Lonzarich[1], and Suchitra E. Sebastian[1]

[1]*Cavendish Laboratory, Cambridge University, JJ Thomson Avenue, Cambridge CB3 OHE, U.K*
[2]*National High Magnetic Field Laboratory, LANL, Los Alamos, NM 87545*
[3]*Physics Department, Northern Illinois University, DeKalb, Illinois 60115*
(Dated: July 22, 2015)

**The normal state in the hole underdoped copper oxide superconductors has proven to be a source of mystery for decades. The measurement of a small Fermi surface by quantum oscillations on suppression of superconductivity by high applied magnetic fields, together with complementary spectroscopic measurements in the hole underdoped copper oxide superconductors, point to a nodal electron pocket from charge order in $YBa_2Cu_3O_{6+\delta}$. Here we report quantum oscillation measurements in the closely related stoichiometric material $YBa_2Cu_4O_8$, which reveal similar Fermi surface properties to $YBa_2Cu_3O_{6+\delta}$, despite an absence of charge order signatures in the same spectroscopic techniques such as x-ray diffraction that revealed signatures of charge order in $YBa_2Cu_3O_{6+\delta}$. Fermi surface reconstruction in $YBa_2Cu_4O_8$ is suggested to occur from magnetic field enhancement of charge order that is rendered fragile in zero magnetic fields because of its potential unconventional symmetry, and/or its occurrence as a subsidiary to more robust underlying electronic correlations.**

**Significance**

Results from quantum oscillation and spectroscopic measurements have suggested charge order as responsible for the creation of a nodal electron pocket in the $YBa_2Cu_3O_{6+\delta}$ and $HgBa_2CuO_{4+\delta}$ families of underdoped cuprates. However, the situation in the pristine $YBa_2Cu_4O_8$ family remains ambiguous. Our high-precision quantum oscillation measurements point to a very similar nodal electron pocket in the stoichiometric cuprate family $YBa_2Cu_4O_8$, despite the non-observation of charge order in diffraction experiments. Our findings in $YBa_2Cu_4O_8$ indicate Fermi surface reconstruction in the hole underdoped cuprates associated with the magnetic field enhancement of charge order, the fragile low-magnetic field character of which is reflected in the short correlation length reported in various materials families.

**Introduction**

The normal state of the underdoped copper oxide superconductors has proven to be even more perplexing than the d-wave superconducting state in these materials. At high temperatures in zero magnetic fields, the normal state of the underdoped cuprates comprises an unconventional Fermi surface of truncated 'Fermi arcs' in momentum space, which is referred to as the pseudogap state [1]. At low temperatures in high magnetic fields, quantum oscillations reveal the non-superconducting ground state in various families of underdoped hole-doped copper-oxide superconductors to comprise small Fermi surface pockets [2–15]. These small Fermi pockets in $YBa_2Cu_3O_{6+\delta}$ have been identified as nodal electron pockets [2, 3, 11, 16, 17] originating from Fermi surface reconstruction associated with charge order measured by x-ray diffraction [18–20], ultrasound [21], nuclear magnetic resonance [22], and optical reflectometry [23]. However, various aspects of the underlying charge order and the associated Fermi surface reconstruction remain obscure. A central question pertains to the origin of this charge order, curious features of which include a short correlation length in zero magnetic field that grows with increasing magnetic field and decreasing temperature [20]. It is crucial to understand the nature of this ground state order that is related to the high temperature pseudogap state and delicately balanced with the superconducting ground state. Here, we shed light on the nature of this state by performing extended magnetic field, temperature, and tilt angle-resolved quantum oscillation experiments in the stoichiometric copper oxide superconductor $YBa_2Cu_4O_8$ [24]. This material with double CuO chains has fixed oxygen stoichiometry, making it a model system to study. $YBa_2Cu_4O_8$ avoids disorder associated with the fractional oxygen stoichiometry in the $YBa_2Cu_3O_{6+\delta}$ chains, which has been shown by microwave conductivity to be the dominant source of weak-limit (Born) scattering [25].

Intriguingly, we find magnetic field- and angle-dependent signatures of quantum oscillations in YBa$_2$Cu$_4$O$_8$ [13, 14] that are very similar to those in YBa$_2$Cu$_3$O$_{6+\delta}$, indicating a similar nodal Fermi surface that arises from Fermi surface reconstruction by charge order with orthogonal wave vectors [16]. However, the same x-ray diffraction measurements that show a Bragg peak characteristic of charge order in YBa$_2$Cu$_3$O$_{6+\delta}$ for a range of hole dopings from $0.084 \leq p \leq 0.164$ [19, 20, 26] hoave, thus far, not revealed a Bragg peak in the case of YBa$_2$Cu$_4$O$_8$ (Fig. 3 in ref. [19]). We suggest that charge order enhanced by applied magnetic fields reconstructs the Fermi surface in YBa$_2$Cu$_4$O$_8$, whereas charge order is revealed even in zero magnetic fields in YBa$_2$Cu$_3$O$_{6+\delta}$ because of pinning by increased disorder from oxygen vacancies.

### Results

Fig. 1 shows quantum oscillations in contactless conductivity [27] measured up to 90 T in YBa$_2$Cu$_4$O$_8$ and at different temperatures from 1.3 K to 8.0 K. The extended magnetic field range and increased sensitivity compared with previous quantum oscillation experiments [13, 14] enable precision measurements of the quantum oscillation frequency spectrum and effective quasiparticle mass of YBa$_2$Cu$_4$O$_8$.

Preliminary quantum oscillation measurements on YBa$_2$Cu$_4$O$_8$ accessed two [13] to four [14] oscillation periods over a restricted magnetic field range, for a magnetic field angle parallel to the crystalline $\hat{c}$-axis. A single quantum oscillation frequency of $660 \pm 30$ T was reported, whereas the scatter of the quantum oscillation amplitude as a function of temperature precluded a determination as to whether a Lifshitz-Kosevich form is obeyed, or an accurate extraction of a quasiparticle effective mass. Our present quantum oscillation measurements over an extended magnetic field range access more than seven oscillation periods, revealing for the first time a pronounced quantum oscillation beat structure characteristic of multiple frequencies, very similar to YBa$_2$Cu$_3$O$_{6.56}$ [6, 9, 28], with a dominant frequency of 640 T. Our precision measurements of quantum oscillation amplitude as a function of temperature shown in Fig. 1b and c further reveal a distinctive Lifshitz-Kosevich form, characteristic of Fermi Dirac statistics. A fit to the Lifshitz-Kosevich form yields a quasiparticle effective mass of $1.8 \pm 1 m_e$ in YBa$_2$Cu$_4$O$_8$, which is, in fact, very similar to that measured for YBa$_2$Cu$_3$O$_{6.56}$.

Angular measurements as a function of tilt angle to the applied magnetic field are required in order to identify the origin of the multiple frequency spectrum that we observe in YBa$_2$Cu$_4$O$_8$. Fig. 2 shows the quantum oscillations in YBa$_2$Cu$_4$O$_8$ measured up to a maximum tilt angle of $\theta \approx 56°$. A few key features are notable. First, the beat pattern at the zero degree tilt angle per-sists up to high tilt angles. Second, the prominent Yamaji amplitude resonance [29] expected at $\theta \approx 52°$ for a neck-belly warped Fermi surface in YBa$_2$Cu$_4$O$_8$ is absent up to the measured high tilt angles. The absence of a prominent Yamaji angle in YBa$_2$Cu$_4$O$_8$ enables us to rule out a Fermi surface where the observed frequency spread arises from a dominant neck and belly fundamental warping. Instead, both of the observed features are consistent with a quasi-two dimensional Fermi surface in which the multiple frequency spectrum originates from a splitting of frequencies as opposed to a fundamental neck and belly warping. Such a quasi-two dimensional Fermi surface is similar to that identified in YBa$_2$Cu$_3$O$_{6+\delta}$ [16, 28].

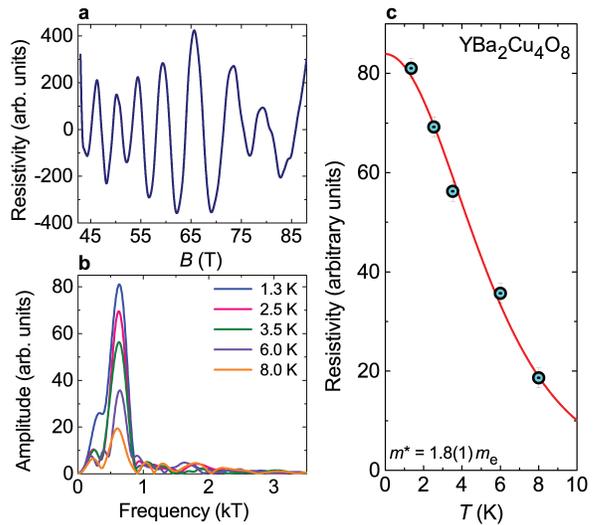

FIG. 1: (a) Quantum oscillations measured in the contactless resistivity of YBa$_2$Cu$_4$O$_8$ for the magnetic field parallel to $c$-axis. (b) Fourier transform of the measured quantum oscillations at different temperatures. (c) Plot of quantum oscillation amplitude as a function of temperature (symbols) accompanied by a Lifshitz-Kosevich fit (red line).

Another clue as to the origin of the observed quantum oscillations in YBa$_2$Cu$_4$O$_8$ is obtained by inspecting the evolution of the amplitude and phase of the quantum oscillations as a function of tilt angle. Fig. 3b shows the cross-correlation function between the measured quantum oscillations in YBa$_2$Cu$_4$O$_8$ and a phase-matched sinusoidal function with frequency 640 T, which is averaged over the indicated magnetic field range, referred to as the correlator. We find that the correlator is very similar to that previously measured for YBa$_2$Cu$_3$O$_{6.56}$ [11, 16] and shown in Fig. 3a. The correlator for YBa$_2$Cu$_4$O$_8$ reveals a zero crossing and phase inversion of the quantum oscillation amplitude, signalling a spin zero at a tilt angle of $\theta \approx 48°$. A spin zero feature arises on destructive interference between quantum oscillations from two spin channels at certain special angles. At these spin zero angles, the ratio of the Zeeman energy to the cyclotron energy – captured by the spin splitting factor $R_s = \cos\left[\frac{\pi}{2}\frac{m^*_\parallel g^*}{m_e \cos\theta}\right]$ – crosses zero, and inverts sign. Spin zero angles are located at

$$\theta_{sz} = \cos^{-1}\left[\frac{m^*_\parallel g^*}{(2n+1)m_e}\right], \quad (1)$$

where $n$ is an integer, $g^*$ is the effective $g$-factor, assumed here to be isotropic for simplicity (see Methods and Table I for the more general case). $m_e$ is the electron mass, and $m^*_\parallel$ is the effective quasiparticle mass for $B$ parallel to the crystalline $\hat{c}$-axis [30].

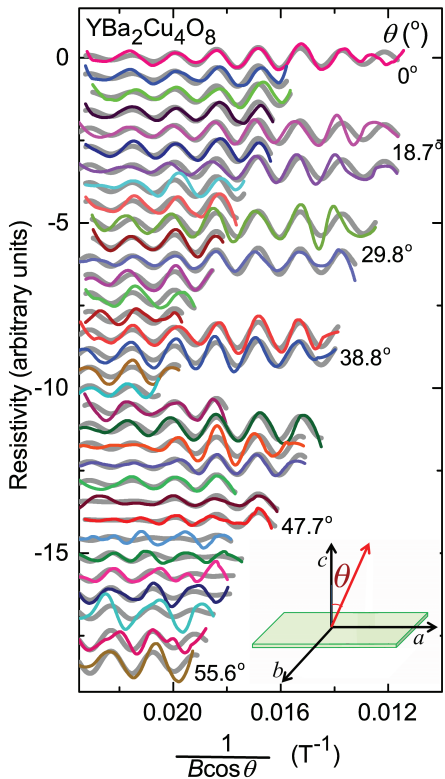

FIG. 2: Quantum oscillations (in colour) measured in the contactless resistivity of $YBa_2Cu_4O_8$ for different angles of inclination ($\theta$) of the magnetic field to the $c$-axis plotted as a function of $\frac{1}{B\cos\theta}$. Simulated quantum oscillations (in gray) of a bilayer split quasi-two dimensional Fermi surface model shown in Fig. 4 [16] are shown for the parameters listed in Table I. The inset shows a schematic of the crystal tilt angle to the magnetic field.

A value of $m^*_\parallel g^*$ can be extracted by the measurement of at least two spin zeros, given the two unknown quantities $n$ and $g^*$, as was done for $YBa_2Cu_3O_{6.56}$ [16]. To interpret the single spin zero observed at $\theta^{sz}_{first} \approx 48°$ in $YBa_2Cu_4O_8$, a comparison with $YBa_2Cu_3O_{6.56}$ is instructive, in which case two spin zeros were observed at $\theta^{sz}_{first} \approx 55°$ [10, 16] and $\theta^{sz}_{sec} \approx 66°$ [16]. Good agreement with experimental data is yielded by associating an index value of $n = 2$ with the first spin zero, for parameters including anisotropic $g$-factors (see Methods) given in Table I. Parameters obtained for $YBa_2Cu_4O_8$ are remarkably similar to those obtained in the case of $YBa_2Cu_3O_{6.56}$ [16], and are in good agreement with experimental data (solid lines in Fig. 3).

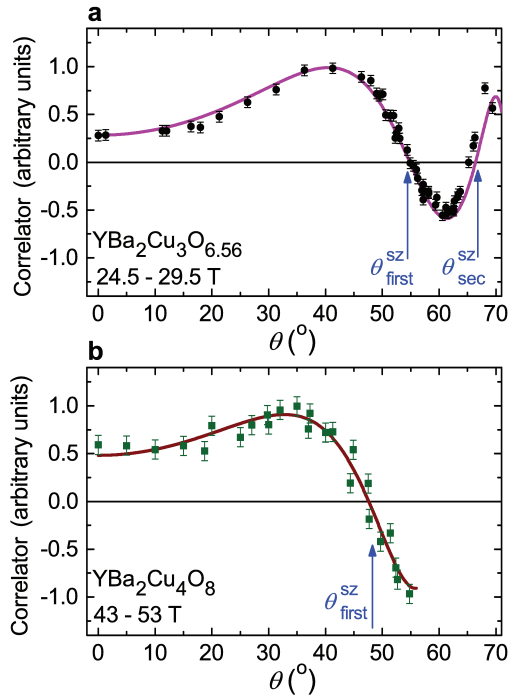

FIG. 3: **a** Symbols represent the cross-correlation function between the measured quantum oscillations in $YBa_2Cu_3O_{6.56}$ and a phase-matched sinusoidal function with frequency 534 T, which is averaged over the indicated magnetic field range (data from ref. [16]). **b** Symbols represent the cross-correlation function between the measured quantum oscillations in $YBa_2Cu_4O_8$ and a phase-matched sinusoidal function with frequency 640 T, which is averaged over the indicated experimental magnetic field range. Solid lines show the fits using the equation and parameters in Table I.

Given the similarities we find between $YBa_2Cu_4O_8$ and $YBa_2Cu_3O_{6+\delta}$ in terms of the measured multiple quantum oscillation frequency spectrum, split quasi-two dimensional Fermi surface, and spin zero angles revealed by the correlator, we compare the measured quantum oscillations in $YBa_2Cu_4O_8$ with the Fermi surface model fit to the measured quantum oscillations in $YBa_2Cu_3O_{6+\delta}$ [11, 16, 17]. We consider a model in which cyclotron orbits are associated with a nodal bilayer split Fermi surface from charge ordering shown in Fig. 4. Here the quantum oscillation frequency spread is associated with a splitting of the Fermi surface arising from tunnelling between bilayers. Magnetic breakdown tunnelling in the nodal region where the splitting is smallest can then give rise to a series of combination frequencies, which is discussed in refs. [16, 28]. Figs. 2 and 3 shows that this model can simulate the angular and magnetic field dependence of the quantum oscillations measured in YBa$_2$Cu$_4$O$_8$ reasonably well. The parameters used for the model simulation are in fact very similar to those used to simulate quantum oscillations measured in YBa$_2$Cu$_3$O$_{6.56}$ (Table I) [16]. The small staggered twofold warping included in the case of YBa$_2$Cu$_3$O$_{6.56}$ is not included for YBa$_2$Cu$_4$O$_8$, given the restricted angular range over which quantum oscillations can be accessed and the likely weaker amplitude warping associated with a longer c-axis. We note that there is some deviation from the model at a few of the highest angles. This deviation may be caused by additional lifting of degeneracy of frequency components from effects such as a distortion of the simple tetragonal crystal structure or other details of Fermi surface geometry beyond those considered in the present model.

| Parameter | Description | YBa$_2$Cu$_4$O$_8$ | YBa$_2$Cu$_3$O$_{6.56}$ |
|---|---|---|---|
| $F_0$ | Quantum oscillation frequency | 639 T | 534 T |
| $\Delta F_{\text{twofold}}$ | Staggered twofold warping frequency | — | 15 T |
| $\Delta F_{\text{split}}$ | Bilayer splitting frequency | 91 T | 90 T |
| $m^*_{\parallel}$ | Quasiparticle effective mass | 1.8 $m_e$ (fixed) | 1.6 $m_e$ (fixed) |
| $B_0$ | Magnetic breakdown field | 4.2 T | 2.7 T |
| $g^*_{\parallel\square}$ | g-Factor 1 | 2.0 | 2.1 |
| $g^*_{\parallel\diamond}$ | g-Factor 2 | 0.1 | 0.4 |
| $\xi_\square$ | g-Factor anisotropy 1 | 1.6 | 1.4 |
| $\xi_\diamond$ | g-Factor anisotropy 2 | 0.8 | 0.2 |
| $\phi$ | Phase | $-1.6^c$ | 0 (fixed) |

TABLE I: Parameters used to simulate the oscillatory waveform for a quasi-two dimensional split Fermi surface model shown in Fig. 4, represented by the Eqn. $\Psi_{\text{twofold}} \approx \sum_{j=1}^{6} N_j [R_{\text{MB}} R_s R_D R_T]_j \cos\left(\frac{2\pi F_j}{B\cos\theta} - \pi + \phi\right)$. Here, $R_{\text{MB}}$ is the magnetic breakdown amplitude reduction factor (defined in Methods), and $N_j$ counts the number of instances that the same orbit is repeated within the magnetic breakdown network. $R_D$ is the Dingle damping factor, $R_T$ is the thermal damping factor, and $R_s$ is the spin damping factor (defined in Methods). The tabulated values used to simulate the quantum oscillation waveform yield good agreement with experiment as a function of $B$ and $\theta$ (Figs. 2 and 3). The effective mass is taken to be a fixed quantity, having been determined independently from temperature-dependent measurements [9]. The parameters are the same for all of the orbits, except for those denoted by subscripts $\square$ and $\diamond$, which each correspond to a subset of orbits as defined in the text. The values of $g^*_{\parallel j}$ and $\xi_j$ here represent parameters used for the simulation rather than unique identifications. The parameters used for YBa$_2$Cu$_3$O$_{6.56}$ shown for comparison are taken from ref. [16].

### Discussion

Our findings therefore reveal closely similar quantum oscillation features in YBa$_2$Cu$_4$O$_8$ compared to YBa$_2$Cu$_3$O$_{6.56}$, showing (i) a similar quasiparticle effective mass from a Lifshitz-Kosevich fit to the amplitude dependence as a function of temperature, (ii) a spread of multiple frequencies yielding a prominent beat structure, (iii) an angular dependence of quantum oscillation frequencies consistent with Fermi surface splitting accompanied by magnetic breakdown rather than fundamen-tal neck and belly warping, and (iv) a similar value of $m^*g^*$ from the angular dependence and spin zero observed in the correlator. The common Fermi surface features point to the same origin of Fermi surface reconstruction in YBa$_2$Cu$_4$O$_8$ as in YBa$_2$Cu$_3$O$_{6+\delta}$, which is particularly relevant given the non-observation of charge order signatures thus far in YBa$_2$Cu$_4$O$_8$. In the case of YBa$_2$Cu$_3$O$_{6+\delta}$, indications are that the electronic structure comprises a nodal electron pocket from Fermi surface reconstruction by a superstructure with orthogonal wave vectors $\mathbf{Q}_1 = 2\pi(\pm\frac{\delta_1}{a}, 0)$ and $\mathbf{Q}_2 = 2\pi(0, \pm\frac{\delta_2}{b})$ associated with charge order observed by x-ray diffraction and other experiments (where $\delta_1 \approx \delta_2 \approx 0.3$) [19, 20, 22].

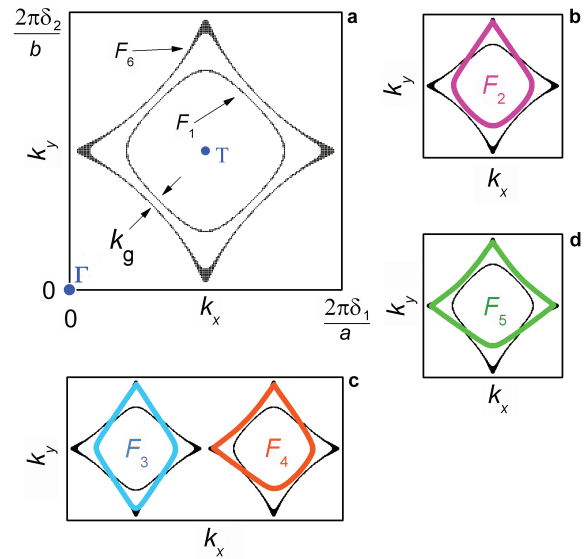

FIG. 4: Brillouin zone cross-section, showing a schematic of a nodal electron pocket created by charge ordering wave vectors $\mathbf{Q}_1 = 2\pi(\pm\frac{\delta_1}{a}, 0)$ and $\mathbf{Q}_2 = 2\pi(0, \pm\frac{\delta_2}{b})$ [11, 16, 17]. **a** shows the two Fermi surface cross-sections of frequency $F_1 = F_0 - 2\Delta F_{\text{split}}$ and $F_6 = F_0 + 2\Delta F_{\text{split}}$. The gap separating bonding and antibonding surfaces is expected to be smallest at the nodes [31]. A cut through the $k_z = 0$ plane of the Brillouin zone shows the possible magnetic breakdown orbits [16, 28]. **b** – **e** show the range of possible magnetic breakdown orbits ($F_2 = F_0 - \Delta F_{\text{split}}$, $F_3 = F_0$, $F_4 = F_0$ and $F_5 = F_0 + \Delta F_{\text{split}}$) as listed in Table I.

An unusual aspect of the charge order measured in YBa$_2$Cu$_3$O$_{6+\delta}$ is the short average correlation length, which has been measured to be of the order of 100 Å in the basal plane and the order of 10 Å along the $c$-axis from x-ray diffraction measurements in a magnetic field of 17 T [20]. The measured average correlation length in the basal plane is comparable with the cyclotron radius of $100 - 200$ Å in the magnetic field range of $30 - 100$ T where quantum oscillations are measured [2–12]. Moreover, given that the anisotropic Fermi velocity associated with the observed pocket has a magnitude along the $c$-axis that is approximately two orders of magnitude lower than in the basal plane [28, 31, 32], the component of the cyclotron trajectories along the $c$-axis is also well within the $c$-axis correlation length. The measured charge ordering wave vectors with finite correlation length inferred from x-ray diffraction experiments represents an average over all parts of the sample, thus representing a lower bound on the correlation length associate with the most highly ordered regions of the sample, which are accessed, for instance, by quantum oscillation measurements. It has remained unclear thus far whether the reported short correlation length is due to the nucleation and pinning of charge order at impurity/defect sites, or whether it is because of the disruptive effects of impurities/defects in an intrinsically long range charge ordered state [33, 34].

Our findings in YBa$_2$Cu$_4$O$_8$, where impurity pinning centres are reduced compared to YBa$_2$Cu$_3$O$_{6+\delta}$ because of the fixed oxygen stoichiometry, suggest that charge order of inherently short correlation length arises in zero magnetic fields in YBa$_2$Cu$_3$O$_{6+\delta}$ on account of impurity sites that act as pinning potential centres. Charge order in the well-known charge density wave material NbSe$_2$ has, for instance, been shown to initially develop around impurities [35]. In the case of YBa$_2$Cu$_4$O$_8$, however, where such impurity pinning centres are reduced, applied magnetic fields are required to tilt the balance of energy scales such that charge susceptibility is further enhanced, and charge order is revealed to reconstruct the Fermi surface.

Our findings point to a fragile form of charge order in the non-superconducting ground state of the underdoped cuprates which is enhanced by applied magnetic fields to yield Fermi surface reconstruction. An interesting question pertains to the origin of such intrinsically fragile charge order. A contributing factor may be the potential unconventional character of the observed charge correlations [36]. Furthermore, charge order may arise as a corollary to more robust correlations in the underdoped cuprates, such as underlying strong spin correlations [37]. The interplay between these correlations may be manifested as other order parameters such as Amperean order [38], pair density wave order [39], d-wave checkerboard order [40, 41], and quadrupolar order [42, 43], which have also been proposed to appear in fluctuating form above the superconducting temperature.

### Methods

Quantum oscillation simulations include conventional thermal, Dingle, and spin damping factors of the same form used for previous comparisons with quantum oscillations measured in the underdoped cuprates and other layered families of materials [30].

The thermal damping factor is given by

$$R_T = \frac{X_j}{\sinh X_j}$$

(where $X_j = 2\pi^2 k_\text{B} m^*_{\theta j} T / \hbar e B$), $k_\text{B}$ is the Boltzmann factor, and $T$ is the temperature; $m^*_{\theta j} = m^*_{\|j}/\cos\theta$ (taken to be the same for all orbits and the subscript '$j$' dropped) is determined by the projection of $B$ perpendicular to the planes (i.e. the projection parallel to the $\hat{c}$-axis in YBa$_2$Cu$_3$O$_{6+x}$), and $m^*_{\|j}$ refers to the value of $m^*_{\theta j}$ when $B$ is parallel to the crystalline $\hat{c}$-axis [taken to be a fixed quantity, determined independently from temperature-dependence measurements [9]].

The Dingle damping factor is given by

$$R_\text{D} = \exp\left(-\frac{\Lambda_j}{B\cos\theta}\right),$$

where $\Lambda_j$ is a damping factor, taken to be the same for all orbits, enabling us to drop the subscript $j$ [30].

The spin damping factor is given by

$$R_\text{s} = \cos\left[\frac{\pi}{2}\left(\frac{m^*_{\theta j}}{m_\text{e}}\right)g^*_{\theta j}\right],$$

where the anisotropic effective $g$-factor has the form $g^*_{\theta j} = g^*_{\|j}\sqrt{\cos^2\theta + \frac{1}{\xi_j}\sin^2\theta}$. Here, $g^*_{\|j}$ refers to the value of $g^*_{\theta_j}$ when $B$ is parallel to the crystalline $\hat{c}$-axis, whereas $\xi_j = \left(\frac{g^*_\|}{g^*_\perp}\right)^2$ is the anisotropy in the spin susceptibility. Two sets of anistropic $g$-factors are considered for subsets of orbits defined below. Because of the multiple frequencies in the model and the restricted angular range measured, it is not possible to uniquely identify the $g$-factors. The

values of $g^*_{\|j}$ and $\xi_j$ here represent parameters used for the simulation.

A splitting of the Fermi surface arising from tunnelling between bilayers leads to two starting frequencies that are denoted as $F_1 = F_0 - 2F_{\text{split}}$ and $F_6 = F_0 + 2F_{\text{split}}$. Magnetic breakdown tunnelling (in the nodal region where the splitting is smallest) gives rise to a series of combination frequencies $F_2, F_3, F_4, F_5$, as discussed in refs. [16, 28]. Only two sets of anisotropic $g$-factors are considered: orbits $F_1, F_2, F_4, F_5$ and $F_6$, which undergo both magnetic breakdown tunnelling and reflection, are approximated to have the same $g$-factor $g^*_{\|\diamond}$ with anisotropy $\xi_\diamond$, whereas orbit $F_3$, which show only magnetic breakdown tunnelling without reflection is approximated to have a common $g$-factor $g^*_{\|\square}$ with anisotropy $\xi_\square$ [16]. The magnetic breakdown amplitude reduction factor is given by

$$R_{\text{MB}} = (i\sqrt{P})^{l_\nu}(\sqrt{1-P})^{l_\eta},$$

in which $l_\nu$ and $l_\eta$ count the number of magnetic breakdown tunnelling and reflection events *en route* around the orbit, having transmitted amplitudes $i\sqrt{P}$ and $\sqrt{1-P}$ respectively. The magnetic breakdown probability is given by $P = \exp(-B_0/B\cos\theta)$, where $B_0$ is the characteristic magnetic breakdown field [30].

### Acknowledgements

We thank B. Keimer, S. A. Kivelson, M. le Tacon, P. A. Lee, C. Pépin, S. Sachdev, and T. Senthil for useful discussions. We also thank the National High Magnetic Field Laboratory personnel, including J. B. Betts, Y. Coulter, M. Gordon, C. H. Mielke, M. D. Pacheco, A. Parish, R. McDonald, D. Rickel, and D. Roybal, for experimental assistance. B.S.T., A.S, and S.E.S. acknowledge support from the Royal Society, the Winton Programme for the Physics of Sustainability, and the European Research Council under the European Unions Seventh Framework Programme (grant number FP/2007-2013)/ERC Grant Agreement number 337425. N.H., Z.Z., F.F.B., and B.J.R. acknowledge support for high-magnetic-field experiments from the US Department of Energy, Office of Science, BES-MSE 'Science of 100 Tesla' programme. G.G.L. acknowledges support from EPSRC grant EP/K012894/1. Work at NIU was supported by the Institute for Nanoscience, Engineering, and Technology. A portion of this work was performed at the National High Magnetic Field Laboratory, which is supported by NSF co-operative agreement number DMR-0654118, the state of Florida, and the DOE.